\documentclass[aps,epsf]{iopart}
\usepackage{graphics}
\usepackage{graphicx}
\usepackage{epsfig}

\begin{document}
\input epsf.tex

\title{Detecting Gravitational Waves from Test-Mass Bodies Orbiting a Kerr Black Hole with P-approximant Templates.}
\author{Edward K. Porter}

\address{Dept. of Physics, Montana State University, Bozeman, 59717, MT, USA.} 

\ead{porter@physics.montana.edu}
\vspace{1cm}
\begin{abstract}
\noindent In this study we apply post-Newtonian (T-approximants) and resummed post-Newtonian (P-approximants) to the case of a 
test-particle in equatorial orbit around a Kerr black hole.
We compare the two approximants by measuring their {\it effectualness} (i.e. larger overlaps with the exact signal), 
and {\it faithfulness} (i.e. smaller biases while measuring the parameters of
the signal) with the exact (numerical) waveforms.  We find that in the 
case of prograde orbits, 
T-approximant templates obtain an effectualness of $\sim 0.99$
for spins $q \leq 0.75.$ For $0.75 < q < 0.95,$ the 
effectualness drops to about 0.82.  The P-approximants achieve effectualness
of $> 0.99$ for all spins up to $q = 0.95.$  The 
bias in the estimation of parameters is much lower in the case of P-approximants
than T-approximants.  We find that P-approximants are both effectual 
and faithful and should be more effective than T-approximants
as a detection template family when $q>0.$
For $q<0$ both T- and P-approximants perform equally well
so that either of them could be used as a detection template family.  However, for parameter estimation, the P-approximant templates still outperforms the T-approximants.
\end{abstract}

\maketitle

\section{Introduction}
Stellar mass compact binaries consisting of double neutron stars (NS),
double black holes (BH) or a mixed binary consisting of a neutron star 
and a black hole, are  the primary targets for a direct first detection of 
gravitational waves (GW) by interferometric detectors, LIGO \cite{LIGO}, VIRGO
\cite{VIRGO},  GEO600 \cite{GEO}, and TAMA~\cite{TAMA}. 
Under radiation reaction the orbit of a binary slowly decays, emitting a signal 
whose amplitude and frequency 
increases with time and is termed a ``chirp'' signal. 
While it is believed that there is a greater population of NS-NS binaries 
\cite{Grish,Phin,Narayan,Stairs,KalogeraAndBelczynski},  
it is the BH-BH binaries that are the strongest candidates for detection since
they can be seen from a greater volume, about two orders-of-magnitude greater
than NS-NS binaries \cite{Grish,PostnovEtAl}.

In order to detect such sources one employs the method of matched 
filtering ~\cite{Helst}.  The effectiveness of matched filtering depends on how well the phase evolution of the waveform is known. Even tiny instantaneous differences,
as low as one part in $10^3$ in the phase of the true signal that
might be present in the detector output and the template that is 
used to dig it out could lead to a cumulative difference of several
radians since one integrates over several hundreds to several thousands
of cycles. In view of improving the signal-to-noise ratio for inspiral
events there has been a world-wide effort in accurately computing the 
dynamics of a compact binary and the waveform it emits or to use
phenomenologically defined detection template families \cite{BCV1,BCV2,BCV3}. 

There have been parallel efforts on using two different approximation schemes: On the
one hand the post-Newtonian (PN) expansion of Einstein's equations has been used
to treat the dynamics of two bodies of comparable masses with and
without spin, in orbit around
each other. This approximation is applicable when the velocities involved 
in the system are small but there is no restriction on the ratio of the
masses \cite{BDIWW,BDI,WillWise,BIWW,Blan1,DJSABF,BFIJ}.  On the other hand, black hole perturbation theory has been used to compute
the dynamics of a test particle in orbit around a spin-less or spinning
black hole. Black hole perturbation theory does not make any assumptions 
on the velocity of the components, but is valid only in the limit when
the mass of one of the bodies is much less than the other
\cite{Poisson1,Cutetal1,TagNak,Sasaki,TagSas,TTS}.  

The post-Newtonian approximation is a perturbative method which expands the equations of 
motion, binding energy and GW flux as a power series in $v/c$, where $v$ is a 
typical velocity in the system and $c$ is the speed of light.  At present, the PN expansion for the case of comparable-masses is known to order 
${\mathcal O}\left(v^{6}\right)$~\cite{DJSABF} and ${\mathcal 
O}\left(v^{7}\right)$~\cite{BFIJ}, for the energy and flux functions,
respectively.  

As previously stated, black hole perturbation theory makes no assumptions about the orbital velocity of the components, but does restrict their masses.  One assumes that 
a test particle of mass $\mu$ is in orbit about a central BH of mass $M$ 
such that $\mu \ll M$.  Assuming this restriction is satisfied we have an analytical 
expression for the energy.  However, no analytical expression has been worked
out for the gravitational wave flux emitted by such a system.  Using black 
hole perturbation theory, the most recent series approximation was calculated 
to ${\mathcal O}\left(v^{11}\right)$ by 
Tagoshi, Tanaka and Sasaki~\cite{TTS}.  For a test particle in circular orbit 
about a Kerr black hole, the most recent progress is a series approximation to 
${\mathcal O}\left(v^{8}\right),$ by Tagoshi, Tanaka, Shibata and 
Sasaki~\cite{SSTT,TSTS}.  

Several authors \cite{TTS,Cutetal2,Poisson3,Poisson4,DIS1} 
have shown that the convergence of both post-Newtonian 
approximation and black hole perturbation theory is too slow 
to be useful in constructing accurate templates.  
More recently,  Damour, Iyer and Sathyaprakash 
(hereafter DIS) showed for the case of a test-mass in orbit about a 
Schwarzschild BH, that by using properly defined energy and flux functions
that have better analytical properties, combined with Pad\'e techniques, it was 
possible to take the existing series expansion and improve its convergence 
properties~\cite{DIS1}.  The new approximation in which Pad\'e approximants
of new energy and flux functions are used to derive improved templates is called
P-approximant.  While in general, more templates are 
needed for P-approximant templates to cover the same volume of parameter 
space~\cite{EKP}, the extra computational cost is preferred
for the increased performance in P-approximants.  

In this paper we will extend the P-approximant technique to the case of 
a test particle orbiting a Kerr black hole.  The reason for focusing on 
test-mass systems is that we can use the exact numerical fluxes \cite{Shibata} from black 
hole perturbation theory with which to compare our results and thereby
reliably demonstrate the usefulness of the technique.

\section{The Gravitational Waveform.} \label{sec:waveform}
In the stationary phase approximation the 
Fourier transform for positive frequencies reads~\cite{Thorne,SathDhur,DWS,DIS2}
\begin{equation}
\tilde{h}(f) \equiv \int_{-\infty}^\infty h(t) \exp(2\pi i f t)\, dt
=\frac{2\eta m {\cal C}}{d} \frac{v^2}{\sqrt{\dot{f}}}
e^{i\left[\psi(f)-\frac{\pi}{4}\right]},
\label{d4.6a}
\end{equation}
where $m=m_{1}+m_{2}$, $\eta=m_{1}m_{2}/m^{2}$, ${\cal C}$ is a constant amplitude coefficient, $d$ is the distance to the source, and, since $h(t)$ is real, $\tilde h(-f) = \tilde h^*(f).$
Also, $v = (\pi m f)^{1/3},$ $\dot f$ is the time-derivative of 
the instantaneous gravitational wave frequency evaluated at the stationary
point given by, 
\begin{equation}
\dot f = -\frac{3v^2}{\pi m^2} \frac{F(v)}{E'(v)},
\end{equation}
where $F$ is the gravitational wave flux function, $E'(v)$ is the derivative of the orbital energy function with respect to $v$, i.e. $E'(v) = dE/dv$, and the evolution of the phase of the 
Fourier transform in the stationary phase approximation is given by solving following set of coupled $1^{st}$ order differential equations 
\begin{equation}
\frac{d\psi}{df} - 2\pi t = 0, \ \ \ \
\frac{dt}{df} + \frac{\pi m^2}{3v^2} \frac{E'(f)}{F(f)} = 0.
\label {eq:fdode}
\end{equation}

\section{Gravitational Binding Energy and Flux Functions}
\label{sec:EnergyAndFlux}
We can see from Equation~(\ref{eq:fdode}) that the phase of the gravitational wave depends both on the energy and flux functions of the binary system.  In the test-mass case we have an exact expression for the energy $E(v)$, but only a series representation for the flux $F(v)$.  Numerically, for the test mass case, the flux has been computed exactly. Since our aim is to draw conclusions on how effective P-approximants are in the comparable
mass case, wherein one has only a Taylor expansion of the flux, we construct
P-approximants of the flux and compare it with the numerical results.
Pad\'e approximation can be thought of as an operator $P_M^N$ 
that acts on a polynomial $\sum_{k=0}^n a_k v^k$ to define a 
rational function $P^{N}_{M}=\sum_{k=0}^{N} A_{k}\,v^{k} / \left(1+\sum_{k=1}^{M} B_{k}\,v^{k}\right)$ such that the $ N + M + 1$ coefficients  in the rational polynomial
on the right hand side is the same as the $n+1$ Taylor coefficients  on the left hand side.
By setting $N=M + \epsilon$ with 
$\epsilon = 0,\ 1,$ we can define two types of Pad\'e approximants: 
These are the super-diagonal, $P^{M+\epsilon}_{M}$, and sub-diagonal, 
$P^{M}_{M+\epsilon}$, approximants.  Normally, the 
sub-diagonal approximants are preferred over super-diagonal approximants.  
 This is because when $M=N+\epsilon$ 
the rational function can be re-expanded as a 
continued fraction which has the property that as 
we go to each new order of the power series only one new coefficient 
needs to be calculated.  Conversely, with the super-diagonal approximants, 
we would have to re-calculate all the $A$'s and $B$'s in the above 
equation as we go to higher orders in the Taylor expansion.  This
means that the sub-diagonal Pad\'e approximants are more {\it stable} and if we
see a trend of convergence in the coefficients the addition of a term
is not likely to spoil this convergence.


\subsection{The Orbital Energy.}\label{subsec:energy}
In the case of both Schwarzschild and Kerr black holes we have an exact 
expression for the orbital energy of a test particle in a circular orbit around 
the parent black hole.  For a black hole of mass $M$ the energy $E$
in terms of the dimensionless magnitude of velocity $v 
\equiv \sqrt{M/r},$ $r$ being is the radial coordinate in the Boyer-Lindquist 
coordinates, takes the form~\cite{Bardeen}
\begin{equation}
E(v, q) = \eta\,\frac{1-2\,v^{2}+q\,v^{3}}{\sqrt{1-3\,v^{2}+2\,q\,v^{3}}}.
\label{eq:energy}
\end{equation}
where $q$ is a dimensionless spin parameter given in terms of the spin
angular momentum $J$ of the black hole by $q \equiv J/M^2 \equiv a/M,$ 
with $a$ spin angular momentum per unit mass in the Kerr metric.


\subsection{Post-Newtonian flux function}

For a test-particle in a circular equatorial orbit, the
post-Newtonian expansion of the flux function has been calculated 
up to ${\mathcal O}$ ($v^{11}$) in the case of a Schwarzschild BH \cite{TTS},
and to ${\mathcal O}$($v^8$) in the case of a Kerr BH \cite{SSTT,TSTS}.  
The general form of the flux function in both these cases is given by the 
expression
\begin{equation}
F_{T_{n}}(x;\, q) = F_{N}(x)\left[\sum_{k=0}^{n}\,a_k(q)x^{k} + 
\ln(x)\,\sum_{k=6}^{n}\,b_{k}(q) x^{k} + {\mathcal O}\left(x^{n+1}\right)\right],
\label{eq:flux}
\end{equation}
where $F_{N}(x)= \frac{32}{5}\eta^{2}x^{10}$ is the dominant {\it Newtonian} flux function. Here, $x$ is the magnitude of the invariant velocity parameter observed 
at infinity which is related to the angular frequency $\Omega$ by $x = 
\left(M\Omega\right)^{1/3}$.  The relation between the parameter $x$ and the 
local linear speed $v$ in Boyer-Lindquist coordinates is given by
\begin{equation}
x(v, q) = v\left[1-qv^{3} + q^{2}v^{6} \right]^{1/3},
\label{eq:xofv}
\end{equation}
which reduces, in the Schwarzschild limit, to $x=v.$ 

\subsection{P-approximant of the flux function}\label{sec:p-approximant}
In order to prepare the series representation of the flux for creating the Pad\'e 
approximation, it is convenient if we factor out the logarithmic terms.  We can 
then write Equation~(\ref{eq:flux}) as
\begin{equation}
F_{T_{n}}(x) = F_{N}(x)
\left[1+\ln\left(\frac{x}{x_{lso}}\right)\sum_{k=6}^{n}\,l_{_{k}}x^{k}\right]
\left[\sum_{k=0}^{n}\,c_{_{k}}x^{k}\right],
\end{equation}
where the new coefficients $c_{k}$ and $l_{k}$ are functions of the old 
coefficients $a_{k}$ and $b_{k}.$ As in Reference~\cite{DIS1} the log-terms 
have been ``normalized'' using the value of the velocity parameter 
at the LSO; this helps in reducing the importance of the log-terms.  $x_{lso}$ is found by substituting $v_{lso}=\sqrt{M/r_{lso}}$ into Equation~(\ref{eq:xofv}), where~\cite{Bardeen}
\begin{equation}
r_{lso}^{\pm}(q) =M\,\left[ 3 + z_{2}(q)\mp \sqrt{\left[ 3 - 
z_{1}(q)\right]\,\left[ 3 + z_{1}(q) + 2\,z_{2}(q) \right]}\right],
\end{equation}
where
\begin{equation}\fl
z_{1}(q)  =  1 + \left(1 - q^2\right)^{\frac{1}{3}}\left[\left(1 + q\right)^{\frac{1}{3}} + \left(1 - 
q \right)^{\frac{1}{3}}\right],\,\,\,\,\,\,\,\,\,\,\,\,\, z_{2}(q)  =  \sqrt{3\,q^2 + z_{1}^{2}},
\end{equation}
and the $+$ $(-)$ sign corresponds to prograde (retrograde) orbits.  We create the factored flux function, $f_{T_{n}}(x)$ by the operation $f_{T_{n}} \equiv \left(1-x/x_{pole}\right)F_{T_{n}}$, where $x_{pole}$ is obtained in the same way as $x_{lso}$ but by using~\cite{Bardeen,Chandra} 
\begin{equation}
r_{pole}^{\pm}(q) = 2\,M \left[ 1 + \cos\left[ \frac{2}{3}\,\cos^{-1}\left(\mp q \right)\right]\right].
\end{equation}
Factoring out the pole also helps to alleviate the problem arising from
the absence of the linear term in the PN expansion of the flux. (Note that
$a_1=0$ in both the Schwarzschild and Kerr cases.)  If we write the expression 
in full we obtain
\begin{equation}
f_{T_{n}}(x) = F_{N}(x)
\left[1+\ln\left(\frac{x}{x_{lso}}\right)\sum_{k=6}^{n}\,l_{_{k}}x^{k}\right]
\left[\sum_{k=0}^{n}\,f_{_{k}}x^{k}\right],
\label{eq:fluxf}
\end{equation}
where $f_{_{0}} = c_{_{0}}$ and $f_{_{k}} = c_{_{k}} - c_{_{k-1}}/x_{pole},$ 
$k = 1,\ldots,n$, 

We can now construct a new flux function by constructing the polynomial expansion
of the inverse of the flux function and construct the Pad\'e approximant of the resulting
polynomial.  We call the approximant constructed this way as {\it  Inverse-} or I-Pad\'e 
approximant because it is obtained from the Taylor expansion of the {\it inverse} 
of the flux function $f(v)$ in Equation~(\ref{eq:fluxf}). The Inverse Pad\'e approximant of flux is defined by
\begin{equation}
f_{IP_{n}}(x) \equiv F_{N}^{-1}(x)
\left[1-\ln\left(\frac{x}{x_{lso}}\right)\sum_{k=6}^{n}\,l_{_{k}}x^{k}\right]P^{
m}_{m+\epsilon}\left[\sum_{k=0}^{n}\,d_{_{k}}x^{k}\right],
\end{equation}
where the coefficients $d_k$ in the Taylor expansion are defined by $\sum_{k=0}^{n}\,d_{_{k}}x^{k} \equiv \left(\sum_{k=0}^{n}\,f_{_{k}}x^{k}\right)^{-1}$.  We call the flux function constructed in this manner \emph{P-approximant}.
Thus, we define the {\it Inverse P-approximant} as,
\begin{equation}
F_{IP_{n}}(x) = 
\left[\left(1-\frac{x}{x_{pole}}\right)\,f_{IP_{n}}(x)\right]^{-1}.
\end{equation}


\section{Effectualness and Faithfulness of T- and P-Approximants}\label{sec:results}
We shall address the performance of the approximants in extracting
the exact waveform in two ways: The {\it effectualness} of the templates
measured in terms of the maximum overlap they can achieve with the exact waveform
when the parameters of the approximant are varied in order to achieve
a good match. The {\it faithfulness} of the approximant templates 
measured in terms of the systematic errors in the estimation 
of parameters while detecting exact waveforms.

\subsection{Overlaps and fitting factor}
We define the scalar product of two waveforms $h$ and $g$ by
\begin{equation}
\left<h\left|g\right.\right> 
=2\int_{0}^{\infty}\frac{df}{S_{h}(f)}\,\left[ \tilde{h}(f)\tilde{g}^{*}(f) +  \tilde{h}^{*}(f)\tilde{g}(f) \right],
\label{eq:scalarprod}
\end{equation}
where the * denotes complex conjugate and 
$\tilde{h}(f),\, \tilde g(f)$ are the Fourier transforms of $h(t),\, g(t)$.
For initial LIGO, the one-sided noise power spectral density (PSD)
from the design study \cite{LIGO} is given by
\cite{DIS2}
\begin{equation}
S_{h}(f) = 9\times10^{-46}\left[ 0.52 + 0.16x^{-4.52} + 0.32 x^{2} \right]\ {\rm Hz}^{-1},
\end{equation}
where $x\equiv f/f_k,$ and $f_{k} = 150$\,Hz is the ``knee-frequency'' of the detector.  
We take the PSD to be infinite below the lower frequency cutoff of $f_{\rm low} = 40$\,Hz.  For two normalized waveforms, or signal vectors, the scalar product returns the 
cosine of the angle between them and is normally referred to as 
the {\it overlap}.  For detection of signals what is more important 
is the {\it fitting factor} $FF:$  As each template is a function 
of the parameters $\lambda^{\mu}$, the fitting factor is defined as 
the maximum overlap obtained by varying the parameters of the template (or the
approximate waveform) relative to the exact waveform:
\begin{equation}
FF = \max_{\lambda^{\mu}} {\mathit O}\left(\lambda^{\mu}\right).
\end{equation}


\begin{figure}[t]
\centerline{\hbox{ \hspace{0.0in} 
    \epsfxsize=2.5in
    \epsffile{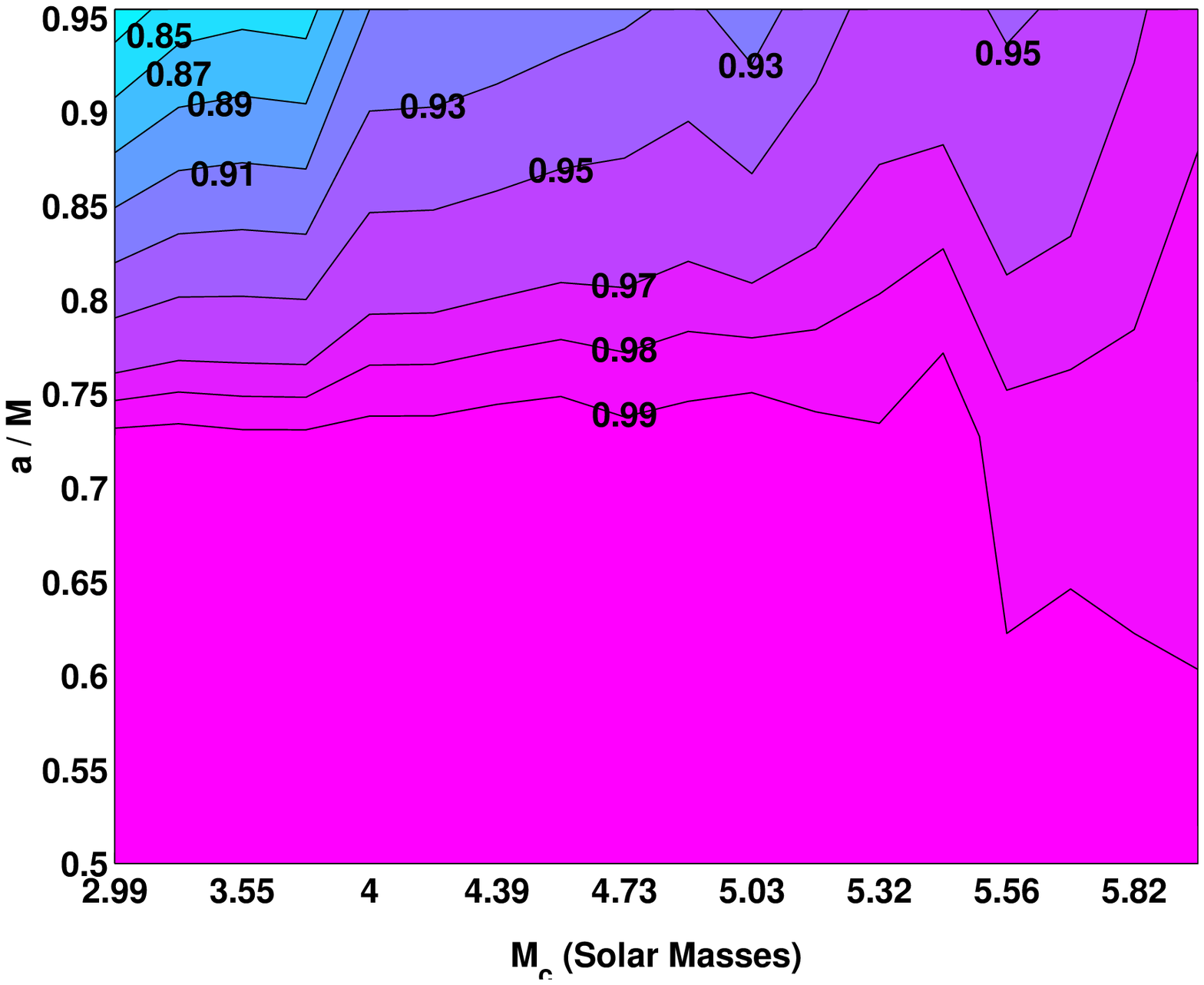}
    \hspace{0.25in}
    \epsfxsize=2.5in
    \epsffile{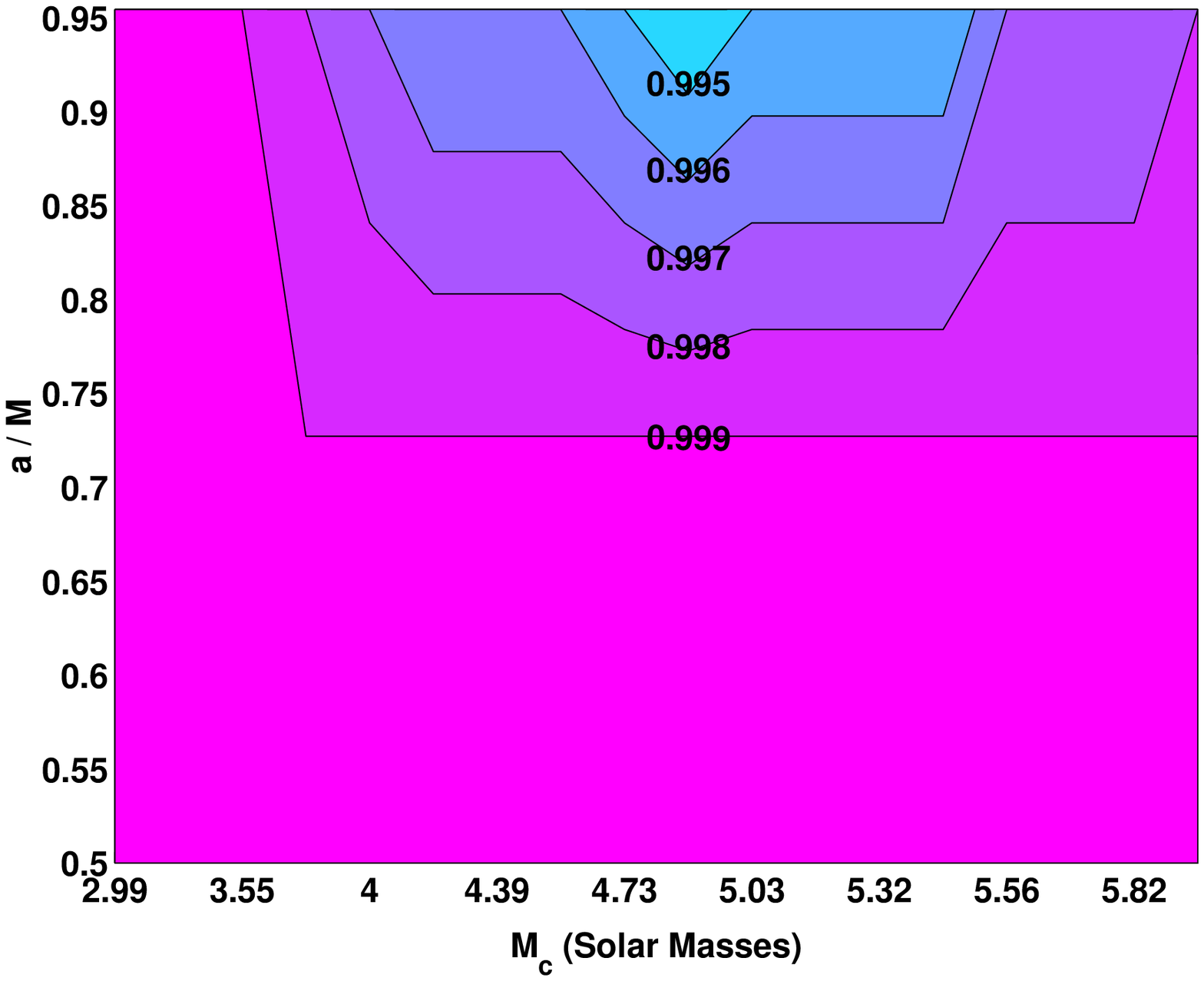}
    }
  }
\caption{The maximized prograde overlaps for T-approximant (left) and
P-approximant (right) templates at the 
$x^{8}$ approximation.  Each system consists of a $1.4\,M_{\odot}$ NS 
inspiralling into a central BH of mass ranging from 10-50 $M_{\odot}$.}
\label{fig:TO}
\end{figure}  
\begin{figure}[t]\vspace{0.2in} 
\centerline{\hbox{ \hspace{0.0in} 
    \epsfxsize=2.3in
    \epsffile{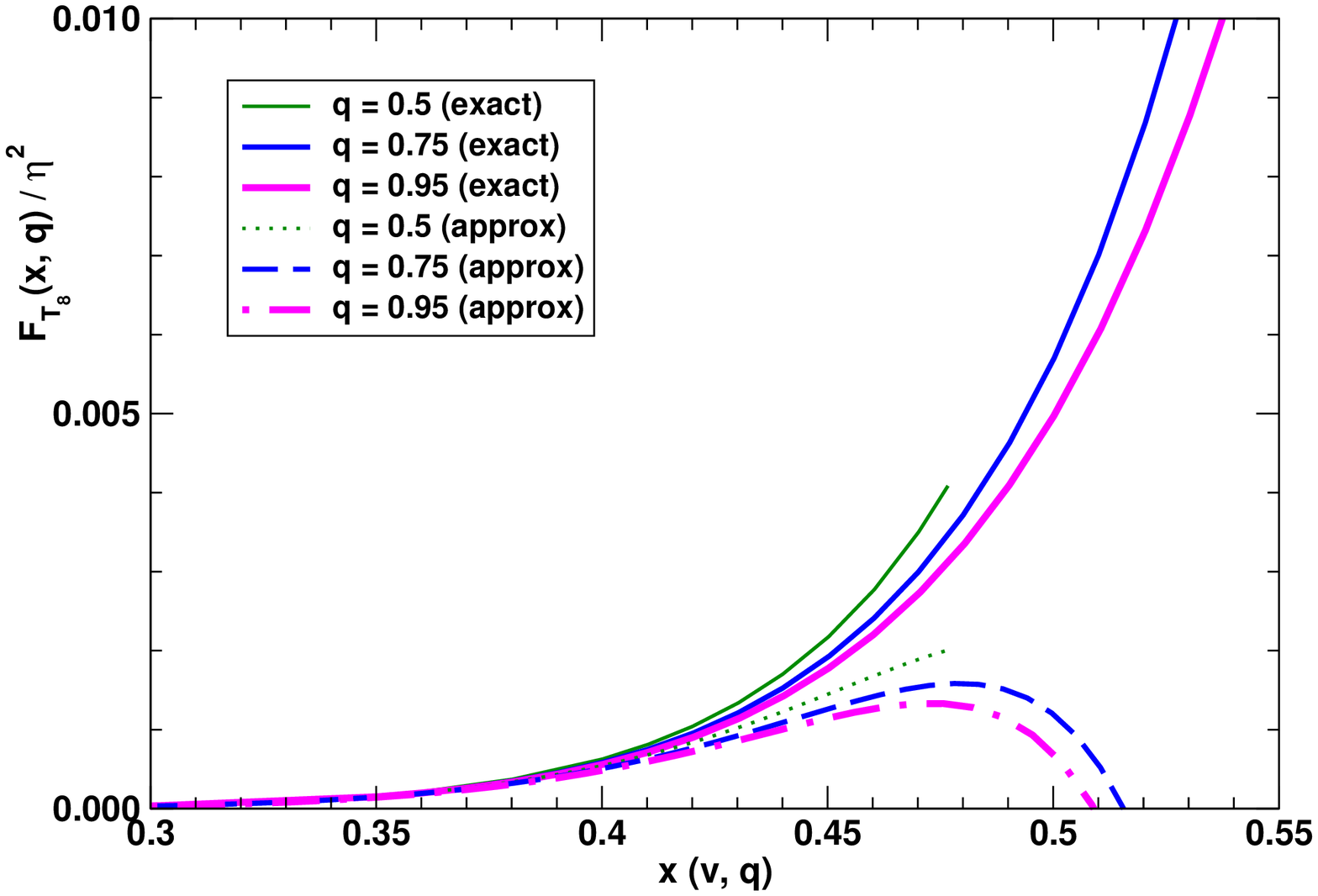}
    \hspace{0.23in}
    \epsfxsize=2.3in
    \epsffile{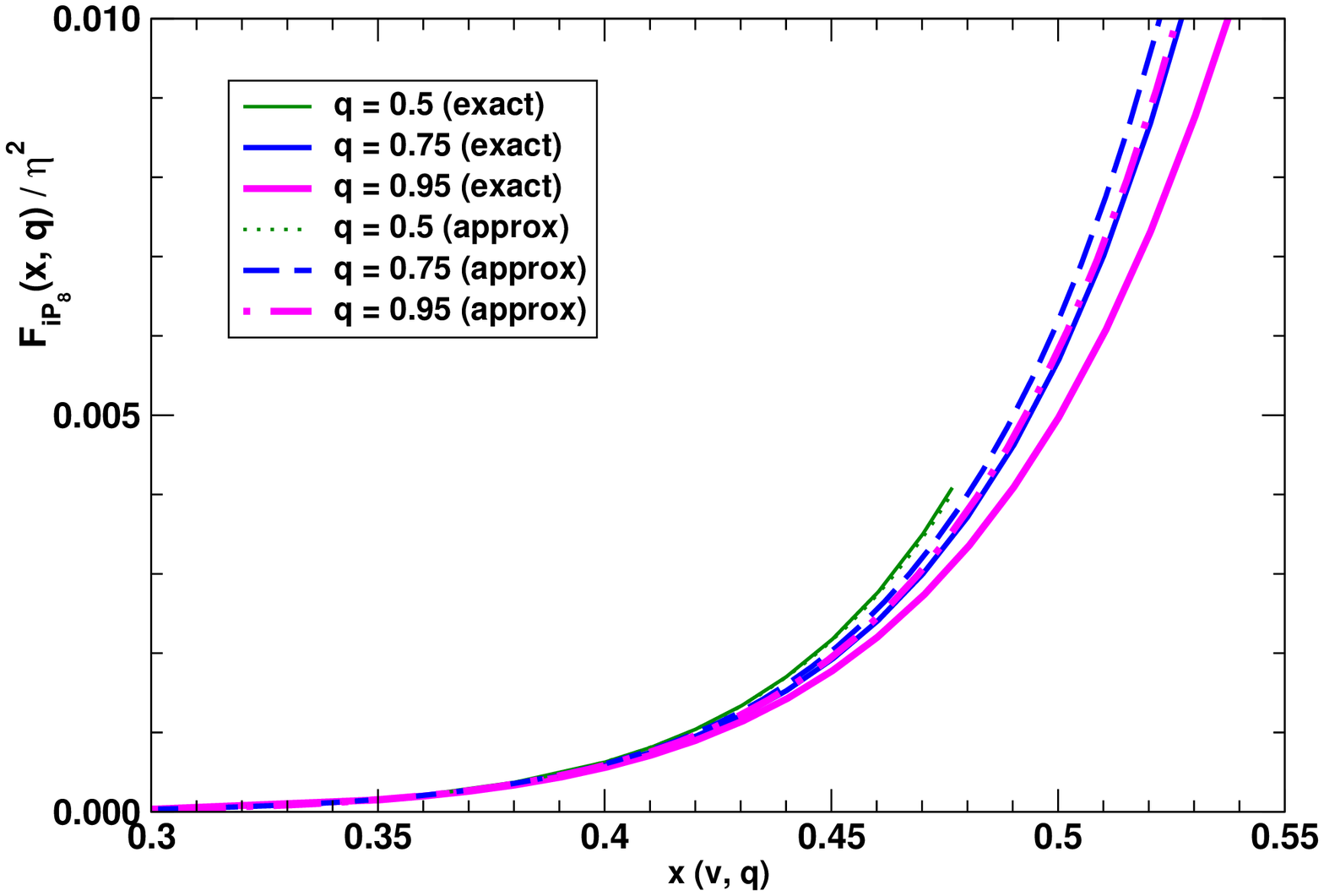}
    }
  }
\caption{The 4-PN T-approximant (left) and P-approximant (right) analytical flux function for spins of $q = 0.5, 0.75$ and $0.95$ against the numerical fluxes for the same spin values.}
\label{fig:x8ff}
\end{figure}  
If two waveforms are a perfect match then their overlap is unity.  In the case of circular equatorial orbits of a test mass around a central black hole, the wave is parameterized by the set $\lambda^{\mu}=(t_{0},\, \Phi_{0},\, m,\, \eta,\, q).$  Maximization over $t_{0}$ is achieved by simply 
computing the correlation of the template with the data in the frequency domain
followed by the inverse Fourier transform. It was pointed out \cite{Schutz2,DhurSath} that the maximum of the overlap of the data with a template
over $\Phi_{0}$ can be computed using just two templates -- an {\it in-phase} and a {\it quadrature-phase} template.   

Using a zero-phase un-normalized waveform, $\tilde{h}_{0} = 
\tilde{h}\left(\Phi_{0} = 0\right)$, we generate two orthonormalized 
waveforms according to $H_{0} = \tilde{h}_{0}/|\tilde{h}_{0}|$ and $H_{\pi /2} =  i H_{0},$ which is explicitly orthogonal to the in-phase template.
The (square of the) maximum of the overlap over $\Phi_{0}$ is given by the
sum-of-squares of the overlap of the signal with the in-phase and quadrature-phase templates:
\begin{equation}
\max_{\Phi_0}\left( {\mathit O} \right) = \sqrt{\left< 
H_{0}\left|h^{\rm X}\right.\right>^{2} + \left<  
H_{\pi/2}\left|h^{\rm X}\right.\right>^{2}},
\end{equation}
where $h^{X}$ denotes the ``exact" waveform.  Once this is done, we use a maximization routine to find the optimal values of the parameters ($m, \eta, q$).

In this study, as we are working in the test-mass approximation, we assume that 
our system is composed of objects with a small mass ratio. For concreteness
we assume that the system comprises a $1.4 M_{\odot}$ NS inspiralling into a central 
BH of varying spin magnitude and mass.  Beginning with a central BH of $10 M_{\odot}$, we 
work our way upward to a $50 M_{\odot}$ BH.  We will also look at the limiting case of a 
$10$-$10 M_{\odot}$ equal-mass system.  Strictly speaking, 
the formulas for energy and flux functions used
in this study are not applicable to the comparable mass case since 
we have neglected the finite mass correction terms in these quantities.
However, the results of such a study should give us an indication
of how strong are the relativistic corrections, as opposed to finite-mass corrections, in the case of comparable masses.  In all cases we vary the spin magnitude of the central BH from $q = -0.95$ to $0.95$. 
Since the main focus in this study is test-mass approximation we shall be
interested in errors in estimating the chirp-mass, ${\mathcal M} = m\eta^{3/5}$, in addition to
the spin magnitude of the central BH.  In view of economy we shall 
only present the results for the highest PN order available, namely
${\cal O}(x^{8})$ order. In all cases our fiducial {\it exact} 
signal $h^{\rm X}$ will be that obtained by using the exact expression for the energy in 
Equation~(\ref{eq:energy}) and the exact numerical fluxes generated using
black hole perturbation theory \cite{Shibata},
and the template will be the approximate waveform constructed using the
exact expression for the energy, as before, and an approximate expression for
the flux.


\begin{figure}
\centerline{\hbox{ \hspace{0.0in} 
    \epsfxsize=2.5in
    \epsffile{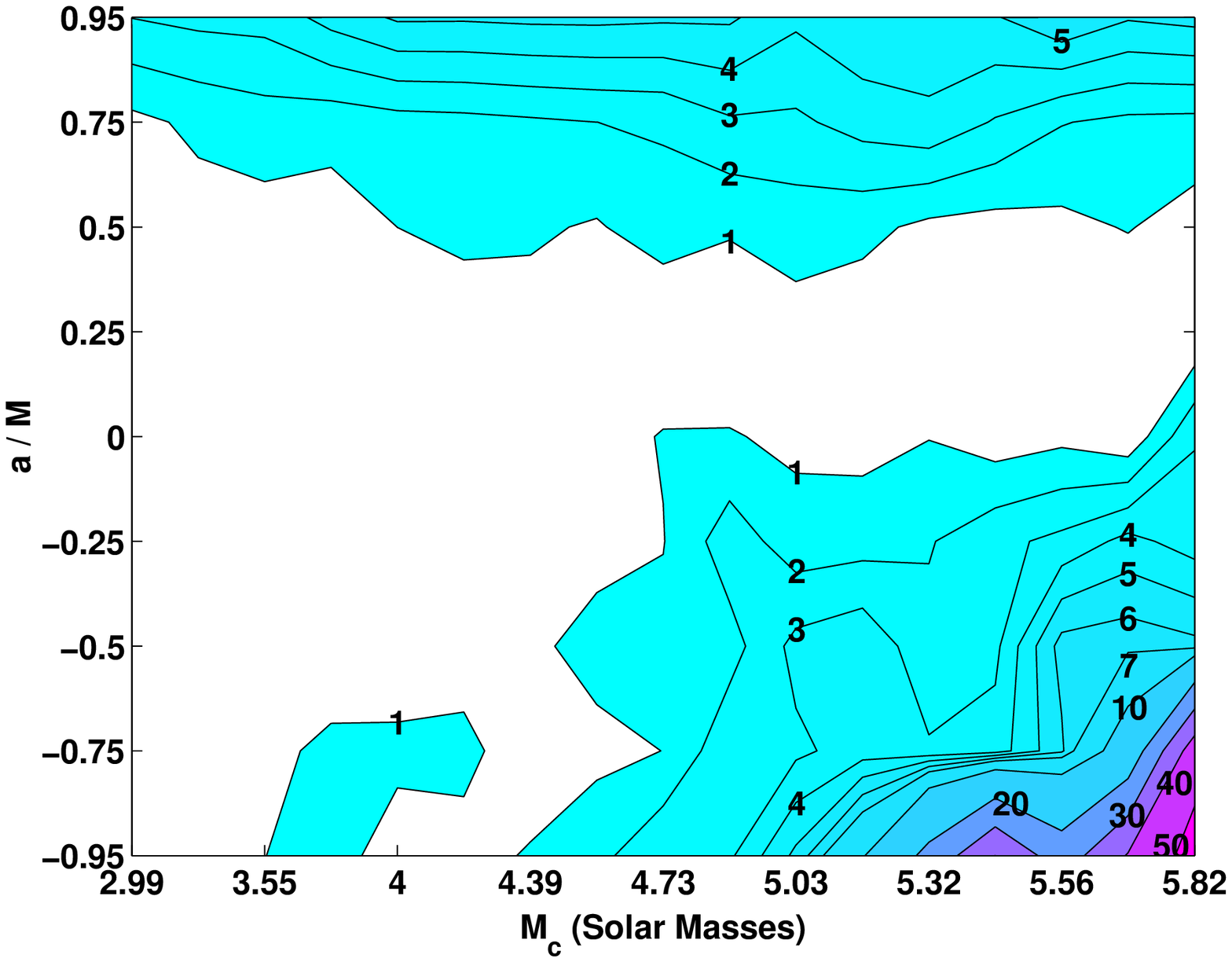}
    \hspace{0.25in}
    \epsfxsize=2.5in
    \epsffile{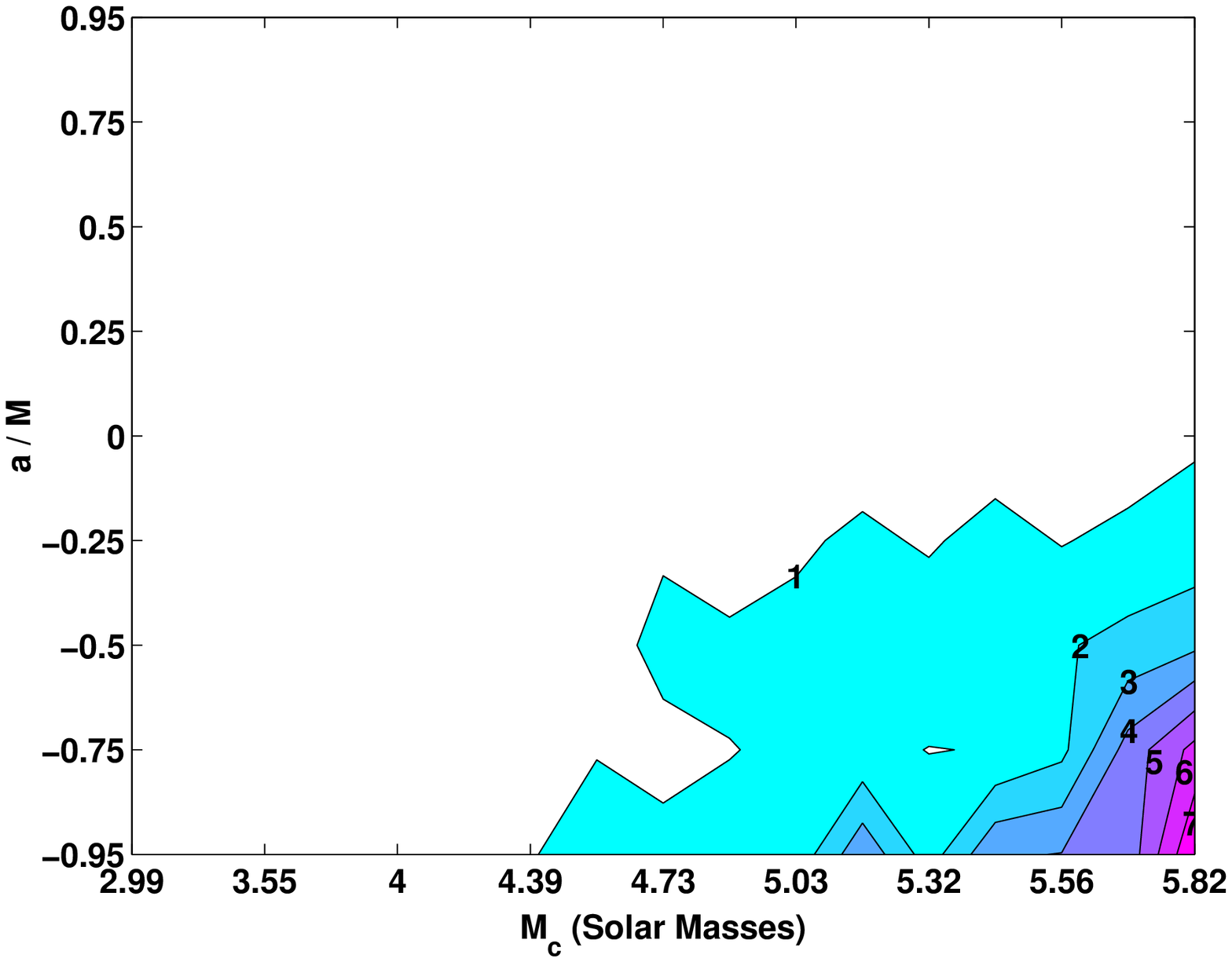}
    }
  }
\caption{The percentage error in the estimation of the chirp-mass, ${\mathcal 
M}$, for T-approximant (left) and P-approximant (right) templates at 
the $x^{8}$ approximation. }
\label{fig:TCME}
\end{figure}  

\subsection{Prograde Orbits -- Effectualness}
\begin{figure}\vspace{11pt}
\begin{center}
\includegraphics[width=3.8in]{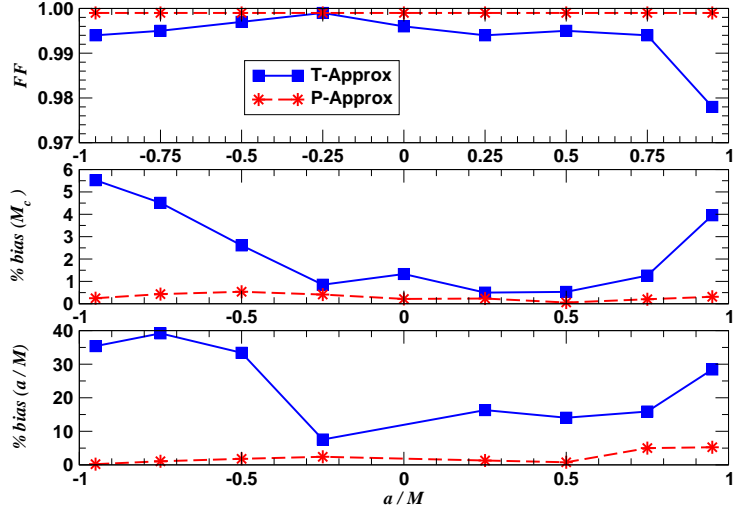}
\caption{The fitting factors for a 10-10 $M_{\odot}$ binary - without the 
finite mass correction terms - for T and P-approximant templates (top).  The 
percentage error in the estimation of the chirp-mass for T and P-approximant 
templates (middle). The percentage error in the estimation of the spin of the 
central black hole for T and P-approximant templates (bottom). }
\label{fig:eqmasspro}
\end{center}
\end{figure}

For T-approximants -- Figure~\ref{fig:TO}, left panel -- maximizing over all parameters gives fitting factors of ${FF \geq 0.98}$ at all mass ranges for the test-mass systems 
up to a spin of $q=0.75$.  Between 0.75 and $q=0.95$, the T-approximant 
templates begin to perform badly and the fitting factors drop to ${FF \sim 
0.82}$.  For the equal-mass case -- Figure~\ref{fig:eqmasspro} --  the templates 
once again achieve fitting factors of ${FF \geq 0.99}$ up to a spin of 
$q=0.75$, but fall off at higher spin magnitudes achieving a fitting factor of 
$\sim 0.98$ at $q = 0.95$.  We should point out that these results do not 
properly convey just how bad the 4-PN T-approximant template actually performs.  
In the left hand panel of Figure~\ref{fig:x8ff}, we plot the PN approximation 
for the flux function against the numerical fluxes at spins of $q = 0.5, 0.75$ 
and $0.95$.  These have corresponding values of $x_{lso} \approx 0.48,0.54$ and $0.66$.  We can see that the flux function at $q = 0.75$ and $0.95$ become 
negative long before the LSO is reached.  This means that as we go to higher and 
higher spins, we can model less and less of the waveform.  
For $q \le 0.6$ we can model the waveform up to the LSO.  However, for $q>0.6$
this we have to stop the waveform generation at a cutoff velocity of 
$x_{cut} \approx 0.5$.  Therefore, the fitting factors beyond $q = 0.75$ completely overestimate the performance of the T-approximant template.

If we now focus on the right hand panel of Figure~\ref{fig:x8ff}, 
we see the true power of the P-approximant templates.  The 4-PN 
template suffers none of the divergences that effect the T-approximants.  
We therefore generate all templates up to the LSO or 2 kHz, whichever 
is reached first.  We can see from the right panel of Figure~\ref{fig:TO} 
that the P-approximant templates achieve fitting factors of $> 0.99$ 
at all spin values.  For the equal-mass case 
-- Figure~\ref{fig:eqmasspro} -- the P-approximant templates achieve fitting 
factors of ${FF \geq 0.995}$ at all mass and spin levels.  This demonstrates 
that in the case of prograde orbits, the P-approximant templates are clearly 
more robust, even at high spin magnitudes of $q=0.95.$

\subsection{Prograde Orbits -- Faithfulness}
In Figure~\ref{fig:TCME} we have 
plotted the percentage bias in the estimation of the chirp-mass for both T-
(left panel) and P-approximant (right panel) templates.  From
Figure~\ref{fig:TCME}, left panel, it is clear that for T-approximants
the bias in ${\mathcal M}$ varies between 0 -- 5$\%.$  
Comparing the left and right panels of Figure~\ref{fig:TCME}
we find that the P-approximant templates (right panels) are more faithful with the 
percentage bias being less than $1\%$ in general.  

\begin{figure}
\centerline{\hbox{ \hspace{0.0in} 
    \epsfxsize=2.5in
    \epsffile{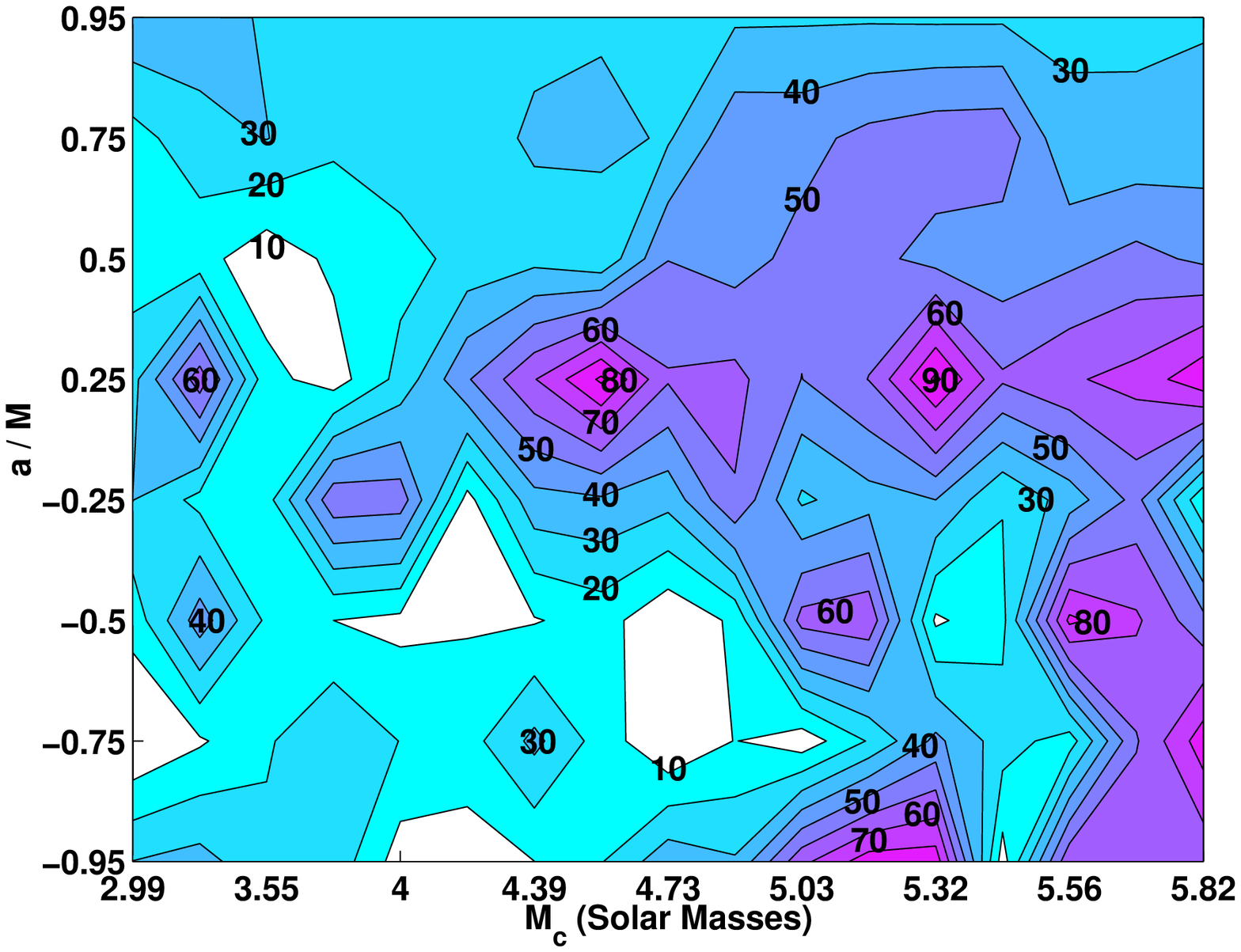}
    \hspace{0.25in}
    \epsfxsize=2.5in
    \epsffile{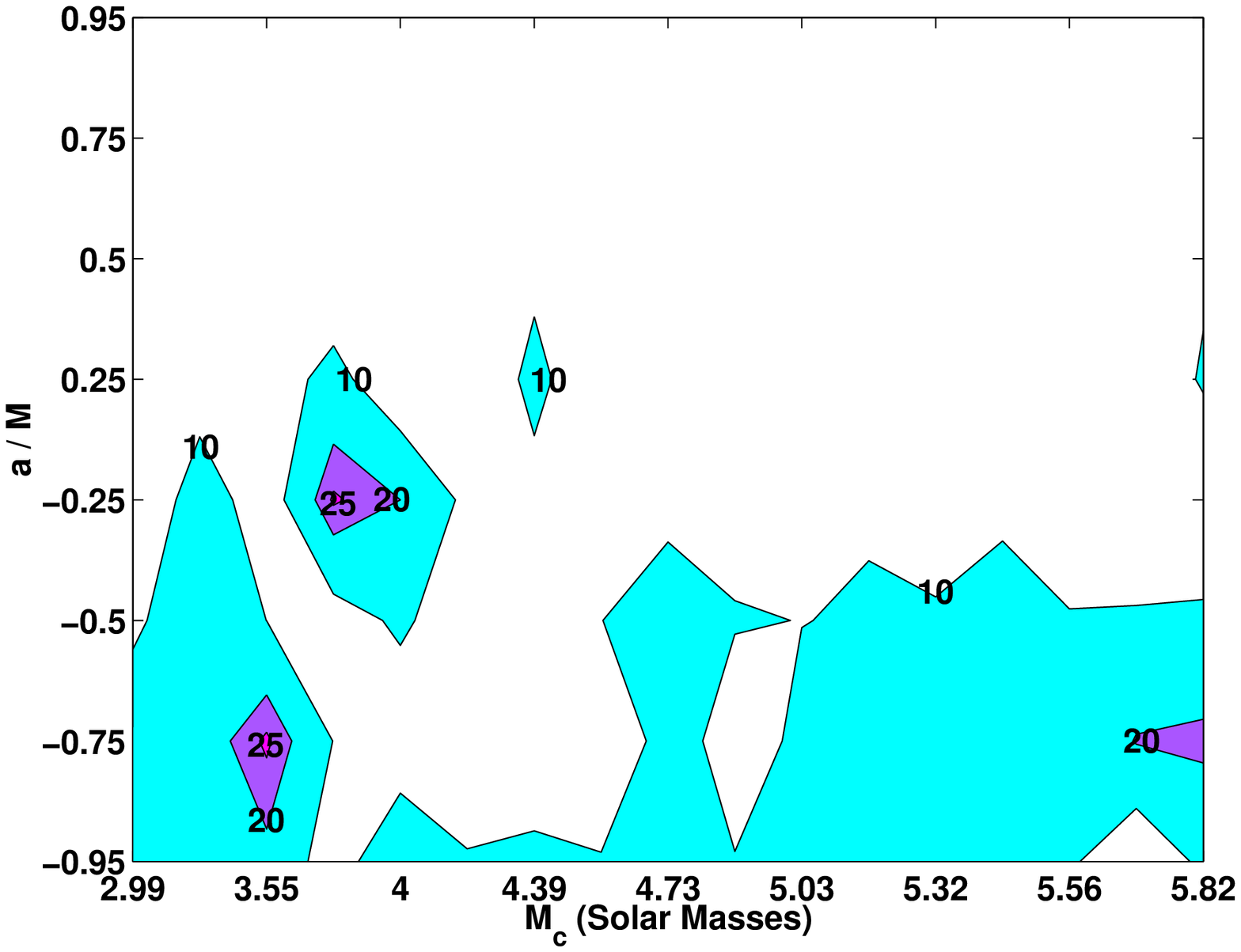}
    }
  }
\caption{The percentage error in the estimation of the spin parameter, $q$, for 
T-approximant (left) and P-approximant (right) 
templates at the $x^{8}$ approximation.}
\label{fig:TAE}
\end{figure}

In Figure~\ref{fig:TAE} we present the percentage bias in the estimation 
of the spin magnitude of the central BH.  We find that for both 
approximants we have to endure a large bias in $q$.  Once again the 
P-approximant templates are more faithful.  For T-approximants the bias 
varies between 5 and 90$\%$, while for P-approximants it lies between 2 and 10$\%$.  
From the middle and bottom Sections of Figure~\ref{fig:eqmasspro}, it is clear that
even in the equal mass case the bias in the estimation of both parameters 
is greater for T-approximant templates than P-approximants.  It is therefore clear that in the case of prograde orbits, P-approximant 
templates must be used in any detection strategy.  


\subsection{Retrograde Orbits -- Effectualness}
In the case of retrograde orbits both templates achieve fitting factors of $FF \geq 0.99$ regardless of 
the spin magnitude and the chirp mass. There is no surprise here: 
The retrograde waveforms are still well within the adiabatic regime and 
are, therefore, modelled well by both templates.  From a purely detection point of 
view, unlike the prograde case, there is no obvious benefit from employing 
P-approximant templates.  For the equal mass system too
there is not much difference in the fitting factors of the two
families of  templates with the exact signal.  

\subsection{Retrograde Orbits -- Faithfulness}
The benefit of using P-approximant templates for retrograde motion is only 
observed when we consider parameter extraction.  Referring to the bottom
portions of the two panels in Figure~\ref{fig:TCME}, we note that the 
T-approximants perform well at all spins for $3.0 \leq {\mathcal M}/M_{\odot} \leq 
4.5$, with a bias of less than $1\%$ in the estimation of ${\mathcal M}$.  
Beyond this, there is in general a bias of $>$ 2$\%$.  As we approach the 
extreme retrograde case the bias rises to as much as 55$\%$.  
The P-approximants perform in a similar manner.  The bias in the 
region $3 \leq {\mathcal M}/M_{\odot} \leq 4.5$ is again in general less than $1\%$.  The 
error does again increase as we head towards the extreme test-mass range, but 
in this case it reaches a maximum value of $8\%$ as opposed to the $55\%$ seen
in the case of  the T-approximants.

The error in estimating $q$ reaches a maximum of $80\%$ for T-approximants and $25\%$ for 
P-approximants.  This is consistent with the results presented in Reference~\cite{SSTT,TSTS} where it was shown that the PN approximation waveforms perform worse in the retrograde case.  Now, referring once more to the equal mass case, the 
P-approximant templates outperform their T-approximant counterparts.
Even with this, we must again conclude that while on the surface there is no 
clear case for using P-approximant templates in searching for retrograde motion 
systems, they should be used because of the lower bias incurred in the estimation of 
parameters.

\section{Conclusions}\label{sec:conclusions}
We have applied P-approximant templates to the case of a NS orbiting Kerr BHs 
of varying mass and spin.  Using a signal waveform constructed from the exact expression for the orbital energy and numerical fluxes from black hole perturbation theory, we were able to compare the performance of T and P-approximant templates.  
In the case of retrograde, Schwarzschild and 
prograde orbits, not only did the P-approximants gave better and more reliable 
fitting factors, they also gave smaller biases in the estimation of parameters.  
We also saw the true power of the P-approximant templates in that we 
were able to generate templates right up to the LSO .  This is something 
that was not possible with the T-approximant templates due to the approximation 
for the flux function becoming negative before the LSO is reached.  
While not being completely correct due to the fact that we omitted the finite-mass correction terms, we also saw that the P-approximant templates gave the best 
performance when in the equal-mass case.  

It is clear that for the type of systems examined in this paper, namely
equatorial test-mass circular orbits in Kerr, P-approximant templates are to be
preferred over their PN counterparts in both detection and measurement for prograde systems.  For retrograde systems, while the T-approximants can be used for detection purposes, the P-approximants are superior when it comes to parameter estimation. 

\section*{Acknowledgments}
The author would like to thank B.S. Sathyaprakash, M. Shibata, T. Damour and B. Iyer.

\end{document}